\documentclass[1p]{elsarticle}

\usepackage{lineno,hyperref}
\usepackage{float}
\usepackage{slashed}
\usepackage{mathtools}
\DeclarePairedDelimiter\bra{\langle}{\rvert}
\DeclarePairedDelimiter\ket{\lvert}{\rangle}
\DeclarePairedDelimiterX\braket[2]{\langle}{\rangle}{#1 \delimsize\vert #2}
\DeclarePairedDelimiterX\inner[2]{\langle}{\rangle}{#1,#2}

\journal{Journal of \LaTeX\ Templates}
\begin{document}

\begin{frontmatter}

\title{$c\bar{b}$ spectrum and decay properties with coupled channel effects}

%% Group authors per affiliation:
%\author{Elsevier\fnref{myfootnote}}
%\address{Radarweg 29, Amsterdam}
%\fntext[myfootnote]{Since 1880.}

%% or include affiliations in footnotes:
\author[mymainaddress]{Antony Prakash Monteiro\corref{mycorrespondingauthor}}
\cortext[mycorrespondingauthor]{Corresponding author}
\ead{aprakashmonteiro@gmail.com}

\author[mymainaddress]{Manjunath Bhat}

\author[mysecondaryaddress]{K. B. Vijaya Kumar}
\address[mymainaddress]{P. G. Department of Physics, St Philomena college
              Darbe, Puttur  574 202, India}
\address[mysecondaryaddress]{Department of Physics, Mangalore University,
Mangalagangothri P.O., Mangalore - 574199, INDIA}

\begin{abstract}
The mass spectrum of $c\bar{b}$ states has been obtained using the phenomenological relativistic quark model
(RQM) with coupled channel effects. The Hamiltonian used in the investigation has  confinement potential and confined one gluon exchange potential (COGEP). In the frame work of RQM a study of M1 and E1 radiative decays of $c\bar{b}$ states has been made. The weak decay widths in the spectator quark approximation have been estimated. An overall agreement is obtained with the experimental masses and decay widths. 
\end{abstract}


\end{frontmatter}

%\linenumbers
\section{INTRODUCTION}

\label{sec:intro}
 The $B_c$ meson is a double heavy quark-antiquark bound state and carries flavours explicitly and provides a good platform for a systematic study of heavy quark dynamics. $B_c$ mesons are predicted by the quark model to be members of the $J^P=0^-$ pseudo scalar ground state multiplet \cite{EC94}. The first successful observation of $B_c$ meson was made by CDF collaboration in 1998 from run I at TEVATRON through the semileptonic decay channel $B_c\rightarrow J/\Psi+l^++\bar{\nu}_l$ \cite{FA98}. They measured the mass of $B_c$ to be $m_{B_c}=6400\pm 390\pm 130$ MeV and the life time $\tau_{B_c}=0.46^{+0.18}_{-0.16}\pm 0.03$ ps. The more precise measurement of mass of $B_c$ i.e.,$m_{B_c}=6275.6\pm 2.9 (\rm stat)\pm 5 (\rm syst)$ $\rm MeV$ was done by the CDF collaboration through the exclusive non-leptonic decay $B_c\rightarrow J/\Psi\pi^+$\cite{AA06,AA706,TAA07}. The results of the CDF collaboration was confirmed by the observations made by the D0 collaboration \cite{VM08,VM09} at TEVATRON. The LHCb has reported several new observations on $B_c$ decays recently. More experimental data on $B_c$ meson are expected in near future from LHCb and TEVATRON.\\

 A suitable theoretical model is required to explain the properties such as mass spectrum, decays, reaction mechanism and bound state behaviour of mesons which involve heavy quarks. The properties of the light and heavy mesons were studied using the phenomenological models. A De Rujula \textit{et al} \cite{DG75} proposed first QCD based model for the study of hadron spectroscopy. The model had a reasonable success and predicted the masses of charmed mesons and baryons. Several non-relativistic phenomenological potentials with radial dependencies for the confinement along with one gluon exchange potential (OGEP) were examined by Bhaduri \textit{et al} \cite{BC81}. The ground state heavy meson spectrum has been studied by Vijaya Kumar \textit{et al} \cite{KB13}. Radiative decay properties of light vector mesons have been studied by Monteiro \textit{et al} \cite{AP14}. Bottomonium spectrum and its decay properties have been studied in a non relativistic model using OGEP by Monteiro \textit{et al} \cite{APM11}. Bhagyesh \textit{et al} \cite{BG12} in their non relativistic model used Hulthen potential to study the orbitally excited quarkonium states. In these models the relativistic effects were completely ignored. \\

There have been many calculations of baryon properties using relativistic models, like MIT bag models \cite{AC74,AR74}, cloudy bag models\cite{AW84,GE79}, chiral bag models \cite{GM79,AG81} etc. Relativistic calculations, where constituent quarks are confined in a potential, have also been performed \cite{SB83,NB86,NS90}. There are other bag models in literature too. In Budapest bag model the volume energy term is replaced by a surface energy term \cite{PJ78}. Another model which effectively contains a surface tension term is the 'SLAC' bag, developed by Bardeen \textit{et al} \cite{WB74} which begins from a local field theory in which heavy quarks interact through a neutral scalar field. Ferreria \textit{et al} \cite{PL77, PL80} used relativistic quark model to study several properties of low lying hadrons. In this model both, the linear and quadratic confinement schemes were used. Bander \textit{et al} \cite{BA84} used a relativistic bound state formalism to make simultaneous study of all meson systems. Isgur \textit{et al} \cite{SN85,SN86} in their relativized quark model used a parametrized potential and incorporated relativistic kinematics to describe all mesons in the same frame work. \\

In NRQM formalism though the mass spectra of ground state $c\bar{b}$ meson has been produced successfully, the radiative decay rates, particularly hindered $M1$ decay rates are significantly influenced by relativistic effects. Therefore, it is necessary to include these effects for the correct description of the decays. Radiative decays are the most sensitive to relativistic effects. Hindered radiative decays are forbidden in the non relativistic limit due to the orthogonality of initial and final meson wave functions. They have decay rates of the same order as the allowed ones. In the relativistic description of mesons an important role is played by the confining quark-antiquark interaction, particularly its Lorentz structure. Thus comparison of theoretical predictions with experimental data can provide valuable information on the form of the confining potential. Hence we use relativistic quark model formalism to study the properties of $c\bar{b}$ meson states.\\

The paper is organized in 4 sections. In sec.~\ref{sec:TB} we briefly review the theoretical background for relativistic model, the framework of the coupled-channel analysis and the relativistic description of radiative decay widths. In sec.~\ref{sec:RD} we discuss the results and the conclusions are drawn in sec.~\ref{sec:C} with a comparison to other models. \\ 

\section{THEORETICAL BACKGROUND}

\label{sec:TB}
\subsection{The Relativistic Harmonic Model}

We investigate properties of $c\bar{b}$ states using confined one gluon exchange potential in the frame work of relativistic harmonic model (RHM) \cite{SB83}. The Hamiltonian used has the confinement potential and a two body confined one gluon exchange potential(COGEP) \cite{PC92, SB91,KB93, KB97}. \\

The confinement potential has Lorentz scalar and a vector harmonic oscillator potential part\cite{VH04,VB09}
\begin{equation}
V_{CONF}(r)=\frac{1}{2}\left(1+\gamma_0\right)A^2r^2+M \label{eq:A}
\end{equation}
where $\gamma_0$ is the Dirac matrix, M is a constant mass and $A^2$ is the confinement strength. \\

 We use the following harmonic oscillator wave equation
\begin{equation}
\left(\frac{p^2}{E+M}+A^2r^2\right)\phi=(E-M)\phi \label{ha}
\end{equation}
the eigenvalue of which is given by
\begin{equation}
E^2_N=M^2+(2N+1)\Omega_N
\end{equation}
where $\Omega_N$ is the energy dependent oscillator size parameter given by
\begin{equation}
\Omega_N=A(E_N+M)^{1/2}
\end{equation}
where $\vec{p}$ is the momentum. For the detailed description of RHM see\cite{SB83,VH04,VB09}.
\subsection{Confined One Gluon Exchange Potential}
 In the present existing models for low energy nuclear phenomena the gluon degrees of freedom have been eliminated from the theoretical frame work and it is assumed that the gluon exchange can be incorporated into the theory through OGEP. But in deriving the OGEP\cite{DG75} the gluon propagators used are similar to the free photon propagators used in obtaining Fermi-Breit interaction in QED. Since the confinement of color means the confinement of quarks as well as gluons, the confined dynamics of gluons should play a decisive role in determining the hadron spectrum and in the hadron-hadron interaction.  The confinement schemes for quarks and gluons have to be more closely connected to each other in QCD and the confinement of gluons has to be taken into account. The COGEP is obtained from the scattering amplitude using confined gluon propagators\cite{PC92,SB91,KB93,KB91}. Here $D^{ab}_{\mu\nu}=\partial^{ab}D_{\mu\nu}$ are the gluon propagators in the momentum representation in current confinement model (CCM)\cite{SB87}. The CCM was developed for the confinement of gluons in the spirit of RHM and aims at a unified confinement theory for the study of quark-gluon bond system  in the spirit of RHM for the confinement of gluons.  In the CCM the coupled non-linear terms in the Yang-Mill tensor is treated as a color gluon super current in analog with Ginzburg-Landu’s theory of superconductivity. The coupled non-linear terms in the equation of motion of  a gluon are simulated by a self induced color current $j_{\mu=\theta^\vartheta_\mu}A_\nu$(=$m^2A_\mu$) or  equivalently an effective mass term for all the gluons with $m^2=c^4r^2-2c^2\delta_{\mu 0}$. The equation of motion is solved using harmonic oscillator modes in the general Lorentz gauge imposing a secondary gauge condition termed ‘oscillator gauge’. The two confined gluon propagators are then obtained in this gauge using the property of the harmonic oscillator wave functions. 
The RHM with COGEP has been quite successful in obtaining  the N-N phase shifts and in hadron spectroscopy\cite{KB93}. \\

The COGEP is obtained from the scattering amplitude \cite{PC92, SB91,KB93, KB97}
\begin{equation}
{\cal M}_{fi}=\frac{g^2_s}{4\pi}\bar{\psi'}_1\gamma^\mu\psi_1D^{ab}_{\mu\nu}(q)\bar{\psi'}_2\gamma^\nu\psi_2 \label{m1}
\end{equation} 
where $\bar{\psi}=\psi^\dagger\gamma_0$, $\psi_{1,2}$ are the wave functions of the quarks in RHM. The $D_{00}(q)~\rm and~D_{ik}(q)$ are the zero energy CCM gluon propagators in momentum representation, where $q=P'_1-P_1=P_2-P'_2$ is the four momentum transfer. $g^2_s/4\pi=\alpha_s$ is the quark gluon coupling constant. In CCM, propagators in the momentum representation are given by,
\begin{equation}
D_{00}(q)=4\pi D_0(q)
\end{equation}
The $D_{ik}(q)$ are given by,
\begin{equation}
D_{ik}(q)=-4\pi\left\lbrace\delta_{ik}-\frac{a^\dagger_{q_i}a_{q_k}}{a_q\cdot a^\dagger_k}\right\rbrace D_1(q)
\end{equation}
where $a_q$ and $a^\dagger_q$ are the creation and destruction operators in the momentum space.\\
The scattering amplitude (\ref{m1}) is written as
\begin{equation}
\begin{split}
{\cal M}_{fi}&=\frac{g^2_s}{4\pi}(\psi'^\dagger_1\psi_1\psi'^\dagger_2\psi_2)D_{00}(q)+\\
&(\psi'^\dagger_1\alpha_i\psi_1)(\psi'^\dagger_2\alpha_k\psi_2)D_{ik}(q)\label{m2}
\end{split}
\end{equation}
We express the 4-component RHM wave function $\psi$ in terms of 2-component wave function $\phi$ by a similarity transformation.\\
i.e.
\begin{eqnarray}
\psi'^\dagger_1\psi_1&=\psi'^\dagger_1U'^\dagger_1(U'^\dagger_1)^{-1}U^{-1}_1U_1\psi_1\\
&=\phi'^\dagger_1(U'^\dagger_1)^{-1}U^{-1}_1\phi_1\label{m3}
\end{eqnarray}
where 
\begin{equation}
N=\sqrt{\frac{2(E+M)}{3E+M}}\label{n1}
\end{equation}
 and 
 \begin{equation}
U=\frac{1}{N\left[1+\frac{p^2}{(E+M)^2}\right]}\left( \begin{array}{cc}
1  &\frac{{\bf \sigma}\cdot{\bf p}}{E+M}\\
-\frac{{\bf \sigma}\cdot{\bf p}}{E+M}&1 \end{array} \right)\label{u1}
\end{equation}
The above expression can be simplified to
\begin{equation}
\psi'^\dagger_1\psi_1=N^2\phi'^\dagger_1\left\lbrace 1+\left[\frac{P^2_1+q\cdot P_1+i\sigma_1\cdot(q\times P_1)}{(E+M)^2}\right]\right\rbrace\phi_1\label{m4}
\end{equation} 
We have,
\begin{equation}
\psi'^\dagger_2\psi_2=\phi'^\dagger_2(U'^\dagger_2)^{-1}U^{-1}_2U_2\phi_2
\end{equation}
i.e.
\begin{equation}
\psi'^\dagger_2\psi_2=N^2\phi'^\dagger_2\left\lbrace 1+\left[\frac{P^2_2+q\cdot P_2+i\sigma_2\cdot(q\times P_2)}{(E+M)^2}\right]\right\rbrace\phi_2\label{m5}
\end{equation} 
Similarly we can write,
\begin{eqnarray}
\psi'^\dagger_1\alpha_i\psi_1=\frac{N^2}{(E+M)}\left[\phi'^\dagger_1\left[2P_1+q+i(\sigma_1\times q)\right]\phi_1\right]_i \label{m6}\\
\psi'^\dagger_2\alpha_k\psi_2=\frac{N^2}{(E+M)}\left[\phi'^\dagger_2\left[2P_2-q-(i\sigma_2\times q)\right]\phi_2\right]_k \label{m7}
\end{eqnarray}
Substituting (\ref{m4}), (\ref{m5}), (\ref{m6}) and (\ref{m7}) in (\ref{m2}), the scattering amplitude now expressed in terms of the two component spinor $\phi$ and the momentum dependent operator $U$ can be written as,
\begin{equation}
{\cal M}_{fi}=4\pi\alpha_sN^4\phi^\dagger_1\phi^\dagger_2\left[U[P_1,P_2,q]\right]\phi_1\phi_2
\end{equation}
The function $U(P_1,P_2,q)$ is the particle interaction operator in the momentum representation and by taking the Fourier transform of each term in the scattering amplitude we get the potential operator $U(\hat{P}_1, \hat{P}_2, r)$ in the co-ordinate space. We drop all the higher order momentum dependent terms in $U(\hat{P}_1, \hat{P}_2, r)$ to obtain the scattering amplitude which is given by
%\begin{widetext}
\begin{equation}
\begin{split}
{\cal M}_{fi}&=4\pi\alpha_sN^4\left[1+\frac{1}{(E+M)^2}\left[\sigma_1\cdot(\nabla\times\hat{P}_1)-\sigma_2\cdot(\nabla\times\hat{P}_2)\right]\right] D_0(\vec{r})+4\pi\alpha_sN^4\\
&\times\left[\frac{1}{(E+M)^2}\left[2\sigma_2\cdot(\nabla\times\hat{P}_1)-2\sigma_1\cdot(\nabla\times\hat{P}_2)-\nabla^2[1-\sigma_1\cdot\sigma_2]-(\sigma_1\cdot\nabla)(\sigma_2\cdot\nabla)\right]D_1(\vec{r})\right]
\end{split}
\end{equation}
%\end{widetext}
The terms which contribute to the central part of COGEP are,\\

~~~~~~~~~~~~~~~~~~~~~~~~~~~~~~~$
D_0(\vec{r})$, $\nabla^2[\sigma_1\cdot\sigma_2-1]D_1(\vec{r})$  and $(\sigma_1\cdot\nabla)(\sigma_2\cdot\nabla)D_1(\vec{r})$\\

In CCM the propagator $D_1(\vec{r})$ satisfies the differential equation
\begin{equation}
(-\nabla^2+c^4r^2)D_1(\vec{r})=4\pi\delta^3(\vec{r})
\end{equation}
The term $(\sigma_1\cdot\nabla)(\sigma_2\cdot\nabla)D_1(\vec{r})$, has angular dependence. But the tensor operator is constructed in such a way that the average value of tensor operator over the angular variables vanishes. The averaging over the direction of r gives
\begin{equation}
(\sigma_1\cdot\nabla)(\sigma_2\cdot\nabla)D_1(\vec{r})=(1/3)\sigma_1\cdot\sigma_2[\nabla^2D_1(\vec{r})]
\end{equation}
Substituting for $[\nabla^2D_1(\vec{r})]$, the central part of the COGEP becomes
%\begin{widetext}
\begin{equation}
V^{cent}_{COGEP}(\vec{r})=\frac{\alpha_sN^4}{4}\vec{\lambda_i}\cdot\vec{\lambda_j}\left[D_0(\vec{r})+\frac{1}{(E+M)^2}\left[4\pi\delta^3(\vec{r})-c^4r^2D_1(\vec{r})\right]\left[1-\frac{2}{3}\vec{\sigma}_i\cdot\vec{\sigma}_j\right]\right]
\end{equation}
%\end{widetext}
where $D_0(\vec{r})$ and $D_1(\vec{r})$ are the propagators given by
\begin{eqnarray}
D_0(\vec{r})=\frac{\Gamma_{1/2}}{4\pi^{3/2}}c(cr)^{-3/2}W_{1/2;-1/4}(c^2r^2)\\
D_1(\vec{r})=\frac{\Gamma_{1/2}}{4\pi^{3/2}}c(cr)^{-3/2}W_{0;-1/4}(c^2r^2)
\end{eqnarray}
where$\lambda_i$ and $\lambda_j$ are color matrices, $\Gamma_{1/2}=\sqrt{\pi}$, W's are Whittaker functions and $c$(fm$^{-1}$) is a constant parameter which gives the range of propagation of gluons and is fitted in the CCM to obtain the glue-ball spectra and r is the distance from the confinement center. 

The terms which contribute to the spin orbit part of the COGEP are
\begin{equation}
[\sigma_1\cdot(\nabla\times\hat{P}_1)-\sigma_2\cdot(\nabla\times\hat{P}_2)]D_0(\vec{r})+
[2\sigma_2\cdot(\nabla\times\hat{P}_1)-2\sigma_1\cdot(\nabla\times\hat{P}_2)]D_1(\vec{r})
\end{equation}
Operating $\nabla$ on $D_0(\vec{r})$ and $D_1(\vec{r})$ and defining
$$\hat{P}=(\hat{P}_1-\hat{P}_2)/2 ~\rm{and}~ \hat{P}_{CM}=\hat{P}_1+\hat{P}_2$$
The spin orbit part of COGEP is
\begin{equation}
\begin{split}
V^{LS}_{12}(\vec{r})=\frac{\alpha_s}{4}\frac{N^4}{(E+M)^2}\frac{\lambda_1\cdot\lambda_2}{2r}
\times\left[[r\times(\hat{P}_1-\hat{P}_2)\cdot(\sigma_1+\sigma_2)](D'_0(\vec{r})+2D'_1(\vec{r}))+\right.\\
\left.[r\times(\hat{P}_1+\hat{P}_2)\cdot(\sigma_1-\sigma_2)](D'_0(\vec{r})-D'_1(\vec{r}))\right]
\end{split}
\end{equation}
The spin orbit term has been split into the symmetric $(\sigma_1+\sigma_2)$ and anti symmetric $(\sigma_1-\sigma_2)$ terms.

The terms which contribute to the tensor part of the COGEP are,
\begin{equation}
\left[(\sigma_1\cdot\nabla)(\sigma_2\cdot\nabla)D_1(\vec{r})-(\frac{1}{3}\sigma_1\cdot\sigma_2[\nabla^2D_1(\vec{r})])\right]
\end{equation}
The tensor part of the COGEP is,
\begin{equation}
V^{TEN}_{12}(\vec{r})=-\frac{\alpha_s}{4}\frac{N^4}{(E+M)^2}\lambda_1\cdot\lambda_2\left[\frac{D''_1(\vec{r})}{3}-\frac{D'_1(\vec{r})}{3r}\right]S_{12}
\end{equation}
where
 \begin{equation}
 S_{12}=[3(\sigma_1\cdot\hat{r})(\sigma_2\cdot\hat{r})-\sigma_1\cdot\sigma_2]
 \end{equation}
\subsection{Coupled Channel Effects}
In this section we briefly review coupled channel models. For detailed discussions on coupled channel effects see \cite{TO79,ON84,TO95,TO96,BE83,KH84,BE80,MI69,Li12,TB97,TS08,ES96,TO85,PG91,HZ91}.

Current QCD inspired potential models generally neglect the hadron loop effects (continuum couplings). These couplings lead to two body strong decays of the meson above threshold and below threshold they give rise to mass shifts of the bare meson states.

In the coupled channel model, the full hadronic state is given by \cite{Li12,TS08,ES96}

\begin{equation}
\ket{\psi}=\ket{A}+\sum_{BC}\ket{BC}
\end{equation}

where we have considered open flavour strong decay $A\to BC$. Here A, B, C denote mesons. 

The wave function $\ket{\psi}$ obeys the equation
\begin{equation}
 H\ket{\psi}=M\ket{\psi}
\end{equation}

The Hamiltonian H for this combined system consists of a valence Hamiltonian $H_0$ and an interaction Hamiltonian $H_I$ which couples the valence and continuum sectors.

\begin{equation}
H=H_0+H_I
\end{equation} 
where 
\begin{equation}
H_I=g\int d^3x\bar{\psi}\psi
\end{equation}

The matrix element of the valence-continuum coupling Hamiltonian is given by \cite{TS08,ES96} 

\begin{equation}
\bra{BC}H_I\ket{A}=h_{fi}\delta (\vec{P}_A-\vec{P}_B-\vec{P}_C)
\end{equation}
where $h_{fi}$ is the decay amplitude.

The mass shift of hadron A due to its continuum coupling to BC can be expressed in terms of partial wave amplitude ${\cal M}_{LS}$ \cite{Li12,ES96}
\begin{eqnarray*}
\Delta M_A^{(BC)}=\int^\infty_0dp\frac{p^2}{E_B+E_C-M_A-i\epsilon}\int d\Omega_p|h_{fi}(p)|^2\\
\quad=\int^\infty_0dp\frac{p^2}{E_B+E_C-M_A-i\epsilon}\sum_{LS}|{\cal M}_{LS}|^2
\end{eqnarray*}
\begin{equation}
\Delta M_A^{(BC)}={\cal P}\int^\infty_0dp\frac{p^2}{E_B+E_C-M_A}\sum_{LS}|{\cal M}_{LS}|^2+i\pi\left(\frac{p*E_B*E_C}{M_A}\sum_{LS}|{\cal M}_{LS}|^2\right)|_{E_B+E_C=M_A}
\end{equation}

The decay amplitude $h_{fi}$ can be combined with relativistic phase space to give the differential decay rate, which is
\begin{equation}
\frac{d\Gamma_{A\to BC}}{d\Omega}=2\pi P\frac{E_BE_C}{M_A}|h_fi|^2
\end{equation}
where in the rest frame of A, we have $\vec{P}_A=0$ and $P=|\vec{P}_B|=|\vec{P}_C|$.
\begin{equation}
P=\sqrt{[M^2_A-(M_B+M_C)^2][M^2_A-(M_B-M_C)^2]}/(2M_A)
\end{equation}
The total decay rate is given by \cite{Li12,ES96}
\begin{equation}
\label{strong}
\Gamma_{A\to BC}=2\pi P\frac{E_BE_C}{M_A}\sum_{LS}|{\cal M}_LS|^2
\end{equation}

\subsection{Radiative Decays} 
Radiative decays are a powerful tool for the study of the quark structure of mesons, and the calculation of corresponding amplitudes is a subject of the increasing interest.
We consider two types of radiative transitions of the $B_c$ meson:

a) Electric dipole (E1) transitions are those transitions  in which the orbital quantum number is changed ($\Delta L = 1$,
$\Delta S = 0$). E1 transitions do not change quark spin. Examples of such transitions are $n ^3S_1\rightarrow n' {^3P_J}\gamma (n > n' )$ and  $n ^3P_J\rightarrow n' {^3S_1}\gamma (n \geq n' )$.
The partial widths for electric dipole (E1) transitions between states $^{2S+1}L_{iJ_i}$ and $^{2S+1}L_{fJ_f}$ are given by
\begin{equation}
\begin{split}
\Gamma_{a\rightarrow b\gamma}=\frac{4\alpha}{9}\mu^2\left(\frac{Q_c}{m_c}-\frac{Q_{\bar{b}}}{m_{\bar{b}}}\right)^2\frac{E_b(k_0)}{m_a}k^3_0\left|\bra{b}r\ket{a}\right|^2\\
\left\{\begin{array}{cl}
(2J+1)/3,& ^3S_1\rightarrow ^3P_J\\
1/3,&^3P_J\rightarrow ^3S_1\\
1/3,& ^1P_1\rightarrow  ^1S_0\\
1,&^1S_0\rightarrow ^1P_1\end{array}\right.
\end{split}
\end{equation}
%\end{widetext}
where $k_0$ is the energy of the emitted photon,\\

~~~~~~~~~~~~~~$k_0=\frac{m^2_a-m^2_b}{2m_a}$ in relativistic model.\\

$\alpha$ is the fine structure constant. $Q_c=2/3$ is the charge of the c quark and $Q_{\bar{b}}=1/3$ is the charge of the $\bar{b}$ quark in units of $|e|$, $\mu$ is reduced mass, $m_{\bar{b}}$ and $m_c$ are the masses of b quark and c quar respectively, $m_a$ and $m_b$ are the masses of initial and final mesons.
\begin{equation*}
\mu=\frac{m_{\bar{b}}m_c}{m_{\bar{b}}+m_c}
\end{equation*}
and\\
$$\frac{E_b(k_0)}{m_a}=1$$
 \begin{equation}
 \bra{b}r\ket{a}=\int^\infty_0 r^3R_b(r)R_a(r)dr
 \end{equation}
 is the radial overlap integral which has the dimension of length, with $R_{a,b}(r)$ being the normalized radial wave functions for the corresponding states.\\
 
 b) Magnetic dipole (M1) transitions are those transitions in which the spin of the quarks is changed ($\Delta S=1, ~~\Delta L=0$) and thus the initial and final states belong to the same orbital excitation but have different spins. Examples of such transitions are vector to pseudo scalar ($n~^3S_1\rightarrow n'~^1S_0+\gamma$, $n\geq n'$) and pseudo scalar to vector ($n~^1S_0\rightarrow n'~^3S_1+\gamma$, $n>n'$) meson decays.\\
The magnetic dipole amplitudes between $S$-wave states are independent of the potential model.

The M1 partial decay width between S wave states is \cite{WJ88,NL78,CE05,WE86,BB95,NB99,BA96}
%\begin{widetext}
\begin{equation}
\label{decay}
\begin{split}
\Gamma_{a\to b\gamma}=\delta_{L_aL_b}4\alpha k^3_0\frac{E_b(k_0)}{m_a}\left(\frac{Q_c}{m_c}+(-1)^{S_a+S_b}\frac{Q_{\bar{b}}}{m_{\bar{b}}}\right)^2(2S_a+1)
\times(2S_b+1)(2J_b+1)\\
\left \{\begin{array}{ccc}
S_a & L_a & J_a \\
J_b & 1  & S_b 
 \end{array} \right\}^2\left \{\begin{array}{ccc}
1 & \frac{1}{2} & \frac{1}{2} \\
\frac{1}{2} & S_a  & S_b 
 \end{array} \right\}^2
\times\left[\int^\infty_0  R_{n_bL_b}(r)r^2j_0(kr/2)R_{n_aL_a}(r) dr\right]^2
\end{split}
\end{equation} 
%\end{widetext}
where $\int^\infty_0 dr  R_{n_bL_b}(r)r^2j_0(kr/2)R_{n_aL_a}(r)$ is the overlap integral for unit operator between the coordinate wave functions of the initial and the final meson states, $j_0(kr/2)$ is the spherical Bessel function. $S_a$, $S_b$, $L_a$, $J_a$ and $J_b$ are the spin quantum number, orbital angular momentum and total angular momentum of initial and final meson states respectively. \\
\subsection{Weak Decays}
\label{weak}
The weak decays of mesons provide information about the underlying quark dynamics within the system. The weak decays of bound state of a quark and an anti-quark,  which carries heavy flavour c and b - enable us to probe the validity of the standard
model of elementary particle interactions and determine several parameters of this model. A rough estimate of the $B_c$ weak decay widths can be done by treating the $\bar{b}$-quark and $c$-quark decay independently so that $B_c$ decays can be divided into three classes \cite{AA99,GS91}$\colon$ (i)the $\bar{b}$-quark decay with spectator $c$-quark, (ii) the $c$-quark decay with spectator $\bar{b}$-quark, and (iii) the annihilation $B^+_c\rightarrow l^+\nu_l$ ($c\bar{s},~u\bar{s}$), where $l=e,~\mu,~\tau$.

\section{Results and Discussion}
\label{sec:RD}
\subsection{Mass Spectrum of $c\bar{b}$ states with coupled-channel effects}

The quark-antiquark wave functions in terms of oscillator wave functions corresponding to the relative and center of mass coordinates have been expressed here, which are of the form, 
\begin{equation}
\Psi_{nlm}(r,\theta, \phi) = N (\frac{r}{b})^{l}~L_{n}^{l+\frac{1}{2}}(\frac {r^2}{b^2})\exp(-\frac{r^2}{2b^2})Y_{lm}(\theta, \phi)
\end{equation}
where N is the normalising constant given by 
\begin{eqnarray}
{\lvert N \rvert} ^2={\frac{2n!}{b^3 \pi^{1/2}}} \frac{2^{[2(n+l)+1]}}{(2n+2l+1)!}(n+l)!
\end{eqnarray} 
$L_{n}^{l+\frac{1}{2}}$ are the associated Laguerre polynomials. \\

The six parameters are the mass of charm quark $m_c$, the mass of beauty quark $m_{\bar{b}}$, the harmonic oscillator size parameter $b$, the confinement strength $A^2$, the CCM parameter $c$ and the quark-gluon coupling constant $\alpha_{s}$. The parameters $m_c, ~m_{\bar{b}},~ A^2$ are obtained by a $\chi^2$ square fit to the available experimental data of charmonium, bottomonium and $B_c$ meson mass spectra. The  CCM parameter $c$ is taken from ref (\cite{PC92,SB85,SB87}) which was obtained by fitting the iota (1440 MeV)$0^{-+}$ as a digluon glue ball.There are several papers in literature where the size parameter $b$ is defined \cite{SN86,IM92}. We obtain the value 'b' by minimizing the expectation value of the Hamiltonian i.e, $\frac{\partial\bra{\psi}H\ket{\psi}}{\partial b}=0$. We then tune the parameter $\alpha_s$ to reproduce the experimental mass value. In literature we find different sets of values for $m_c$ and $m_{\bar{b}}$, which are listed in Table \ref{mass1}.
\begin{table}[!h]
\centering
\caption{\label{mass1}\bf m$_c$ and m$_{\bar{b}}$ for various theoretical models (in MeV).}
\setlength{\tabcolsep}{2pt}
\begin{tabular}{ccccccc}
\hline
Parameter&Ref.\cite{AM80}&Ref. \cite{EJ78}&Ref. \cite{WS81}& Ref. \cite{CJ77}& Ref.\cite{DR03}\\
\hline
$m_c$&1800&1480 &1480&1480 &1550 \\

$m_{\bar{b}}$&5174 &5180 &4880 &4880 &4880 \\
\hline
\end{tabular}
\end{table}
The values of strong coupling constant $\alpha_s$ in literature are listed in Table \ref{alpha}. The value of strong coupling constant ($\alpha_{s}$=0.3) used in our model is compatible with the perturbative treatment.\\
\begin{table}[!h]
\centering
\caption{\bf $\alpha_s$ for various theoretical models.}
\label{alpha}
\setlength{\tabcolsep}{2pt}
\begin{tabular}{ccccccc}
\hline
Parameter&Ref. \cite{SN85}&Ref. \cite{DR03}&Ref.\cite{AA05}& Ref. \cite{EC94}& Ref. \cite{SA95}\\
\hline
$\alpha_s$& 0.21&0.265&0.357&0.361&0.391\\
\hline
\end{tabular}
\end{table}
We use the following set of parameter values.
\begin{equation}
\begin{split}
&m_c = 1525.00\pm 0.37~~~{\rm MeV};~~~m_{\bar{b}}=4825.00\pm 0.29~{\rm MeV};\\
&b
= 0.3~{\rm fm};~~~ \alpha_s = 0.3;~~A^2=550.00 \pm 0.78~{\rm MeV~fm^{-2}};c=1.74~{\rm fm^{-1}}
\end{split}
\end{equation}
\begin{table*}[!h]
\centering
\caption{\label{Shift}\bf Mass shifts (in MeV).}
\begin{tabular*}{\textwidth}{@{\extracolsep{\fill}}lrrrrrrrrl@{}}
\hline

 &&\\
Bare $c\bar{b}$ State&\multicolumn{1}{c}{BD}&\multicolumn{1}{c}{$B_sD_s$}&\multicolumn{1}{c}{$B_0D_0$}&\multicolumn{1}{c}{$B^*D$}&\multicolumn{1}{c}{$B^*_sD_s$}&\multicolumn{1}{c}{$B^*D^*$}&\multicolumn{1}{c}{$B^*_sD^*_s$}&\multicolumn{1}{c}{Total}\\
\hline
$1~^1S_0$&0&0&0&-5.661&-5.033&-10.434&-9.328&-30.456\\

$1~^3S_1$&-2.046&-1.805&-2.052&-3.955&-3.496&-7.293&-6.488&-27.135\\
$1~^3P_0$&-57.922&-57.406&-57.946&0&0&-19.088&-18.932&-211.294\\
$1~^1P_1$&0&0&0&-18.49&-18.393&-37.603&-37.901&-112.387\\
$1~^3P_1$&0&0&0&-38.390&-38.049&0&0&-76.439\\
$1~^3P_2$&-40.618&-40.314&-40.632&0&0&0&0&-121.557\\
$2~^1S_0$&0&0&0&-1.547&-1.361&-2.837&-2.523&-8.268\\
$2~^3S_1$&-0.546&-0.476&-0.548&-1.929&-1.711&-1.049&-0.920&-7.179\\
$1~^3D_1$&-30.675&-30.312&-30.682&-15.326&-15.146&-3.077&-3.044&-128.262\\
$1~^1D_2$&0&0&0&-3.147&-3.111&-27.643&-27.49&-61.391\\
$1~^3D_2$&0&0&0&-27.214&-27.552&-69.486&-68.957&-193.209\\
$1~^3D_3$&-40.753&-40.359&-40.772&-54.308&-53.783&-20.835&-20.606&-230.663\\
$2~^3P_0$&-148.72&-146.395&-148.828&0&0&-48.589&-47.903&-540.435\\
$2~^1P_1$&0&0&0&-25.081&-24.744&-49.343&-48.741&-147.909\\
$2~^3P_1$&0&0&0&-98.623&-97.088&0&0&-195.711\\
$2~^3P_2$&-79.114&-77.890&-79.171&0&0&0&0&-236.175\\

\hline
\end{tabular*}
\end{table*} 

\begin{table*}[!h]
\centering
\caption{\label{spectrum}\bf $B_c$ meson mass spectrum (in MeV).}
\begin{tabular*}{\textwidth}{@{\extracolsep{\fill}}lrrrrrrrrl@{}}
\hline
State &&\\
$n~^{2S+1}L_J$&\multicolumn{1}{c}{This work}&\multicolumn{1}{c}{Ref.\cite{SJ96}}&\multicolumn{1}{c}{Ref. \cite{VA95}}&\multicolumn{1}{c}{Ref. \cite{ZV95}}&\multicolumn{1}{c}{Ref. \cite{EC94}}&\multicolumn{1}{c}{Ref.\cite{DR03}}&\multicolumn{1}{c}{Ref.\cite{SN85}}&\multicolumn{1}{c}{Ref.\cite{CT96}}&\multicolumn{1}{c}{Ref.\cite{FL}}\\
\hline
$1~^1S_0$&6275&6247&6253&6260&6264&6270&6271&6280$\pm 30\pm 190$&6286\\

$1~^3S_1$&6314&6308&6317&6340&6337&6332&6338&6321$\pm 20$&6341\\
$1~^3P_0$&6672&6689&6683&6680&6700&6699&6706&6727$\pm 30$&6701\\
$1P1$&6766&6738&6717&6730&6730&6734&6741&6743$\pm 30$&6737\\
$1P1'$&6828&6757&6729&6740&6736&6749&6750&6765$\pm 30$&6760\\
$1~^3P_2$&6776&6773&6743&6760&6747&6762&6768&6783$\pm 30$&6772\\
$2~^1S_0$&6838&6853&6867&6850&6856&6835&6855&6960$\pm 80\pm$&6882\\
$2~^3S_1$&6850&6886&6902&6900&6899&6881&6887&6990$\pm 80$&6914\\
$1~^3D_1$&7078&&7008&7010&7012&7072&7028&&7019\\
$1D2$&7009&&7001&7020&7012&7077&7041&&7028\\
$1D2'$&7154&&7016&7030&7009&7079&7036&&7028\\
$1~^3D_3$&6980&&7007&7040&7005&7081&7045&&7032\\
$2~^3P_0$&6914&&7088&7100&7108&7091&7122&&\\
$2P1$&7259&&7113&7140&7135&7126&7145&&\\
$2P1'$&7322&&7124&7150&7142&7145&7150&&\\
$2~^3P_2$&7232&&7134&7160&7153&7156&7164&&\\

\hline
\end{tabular*}
\end{table*} 
  We evaluate the bare state masses and shifts due to $BD$,$B_sD_s$, $B^0D^0$, $B^*D$, $B^*_sD_s$, $B^*D^*$ and $B^*_sD^*_s$ loops (with $M_B=5279.26 ~\rm{MeV}$, $M_{B_s}=5366.77~ \rm{MeV}$, $M_{B^0}=5279.58~ \rm{MeV}$, $M_{B^*}=5324.6 ~\rm{MeV}$, $M_{B^*_s}=5415.4~ \rm{MeV}$, $M_{D}=1869.61~ \rm{MeV}$, $M_{D_s}=1968.30~ \rm{MeV}$, $M_{D_0}=1864.84~\rm{MeV}$, $M_{D^*}=2006.96~\rm{MeV}$ and $M_{D^*_s}=2112.1$).\\
  
   For the case of a bound system of quark and antiquark of unequal mass, charge conjugation parity is no longer a good quantum number so that states with different total spins but with the same total angular momentum, such as the $^3P_1 - ^1P_1$ and $^3D_2 - ^1D_2$ pairs, can mix via the spin orbit interaction or some other mechanism. The $B_c$ meson states with $J=L$ are linear combination of spin triplet $\ket{^3L_J}$ and spin singlet $\ket{^1L_J}$ states which we describe by the following mixing$\colon$
 \begin{eqnarray}
&\ket{nL'}=\ket{n~^1L_J}\cos\theta_{nL}+\ket{n~^3L_J}\sin\theta_{nL}\\
&\ket{nL}=-\ket{n~^1L_J}\sin\theta_{nL}+\ket{n~^3L_J}\cos\theta_{nL}
\end{eqnarray}
$$~J=L=1,2,3,\cdots$$
where $\theta_{nL}$ is a mixing angle, and the primed state has the heavier mass. For $L=J=1$ we have mixing of P states 
  \begin{eqnarray}
&\ket{nP'}=\ket{n~^1P_1}\cos\theta_{nP}+\ket{n~^3P_1}\sin\theta_{nP}\\
&\ket{nP}=-\ket{n~^1P_1}\sin\theta_{nP}+\ket{n~^3P_1}\cos\theta_{nP}
\end{eqnarray}
The values of the mixing angles for P states are $\theta_{1P}=0.4^\circ$ and $\theta_{2P}=0.05^\circ$\\

Similarly for $L=J=2$ we have mixing of D states,
\begin{eqnarray}
&\ket{nD'}=\ket{n~^1D_2}\cos\theta_{nD}+\ket{n~^3D_2}\sin\theta_{nD}\\
&\ket{nD}=-\ket{n~^1D_2}\sin\theta_{nD}+\ket{n~^3D_2}\cos\theta_{nD}
\end{eqnarray}
The value of mixing angle for D states is
$\theta_{1D}=0.05^\circ$\\
 The calculated masses of the $c\bar{b}$ states are listed in Table ~\ref{spectrum}. Our calculated mass value for $B_c(1S)$ is 6275.851 MeV and for $B^*_c$(1S) is 6314.161 MeV. $B^*_c$(1S) is heavier than $B_c$(1S) by 38.193 MeV. This difference is justified by calculating the $^3S_1-{}^1S_0$ splitting of the ground state which is given by 
 \begin{equation}
 M({}^3S_1)-M({}^1S_0)=\frac{32\pi\alpha_s|\psi(0)|^2}{9m_cm_b}
 \end{equation}
 The mass of first radial excitation $B_c$(2S) is 6838.232 MeV which is heavier than $B_c$(1S) by 562.381 MeV. This value agrees with the experimental value of $B_c$(2S) 6842$\pm$4$\pm$5 MeV\cite{AG14}. The difference between the $B^*_c$(2S) and $B^*_c$(1S) masses turns out to be 536.412 MeV.  Our prediction for masses of orbitally excited $c\bar{b}$ states are in good agreement with the other model calculations.

\subsection{Radiative Decays}
The calculation of radiative (EM) transitions between the meson states can be performed from first principles in lattice QCD, but these calculation techniques are still in their developmental stage. At present, the potential model approaches provide the detailed predictions that can be compared to experimental results. \\

The possible $E1$ decay modes have been listed in Table \ref{E1} and the predictions for E1 decay widths are given. Also our predictions have been compared  with other theoretical models. Most of the predictions for $E1$ transitions are in qualitative agreement. However, there are some differences in the predictions  due to differences in phase space arising from different mass predictions and also from the wave function effects. For the transitions involving $P1$ and $P1'$ states which are mixtures of the spin singlet $^1P_1$ and spin triplet $^3P_1$ states, there exists huge difference between the different theoretical predictions. These may be due to the different $^3P_1 - ^1P_1$ mixing angles predicted by the different models. Wave function effects also appear in decays from radially excited states to ground state mesons such as $2~^3P_0\rightarrow 1~^3S_1\gamma$. The overlap integral for these transitions in our model vanishes and hence we get decay width for these transitions zero.\\

 The M1 transitions contribute little to the total widths of the 2S levels. Because it cannot decay by annihilation.
Allowed M1 transitions correspond to triplet-singlet transitions between S-wave states of the same n quantum number, while hindered M1 transitions are either triplet-singlet or singlet-triplet transitions between S-wave states of different n quantum numbers.\\

 The possible radiative M1 transition modes are as follows, (i) 2 $^3S_1\rightarrow 2 ^1S_0+\gamma$, (ii) 2 $^3S_1\rightarrow 1 ^1S_0+\gamma$, (iii) 2 $^1S_0\rightarrow 1 ^3S_1+\gamma$, (iv)1 $^3S_1\rightarrow 1 ^1S_0+\gamma$. 
In the above (ii) and (iii) represent hindered transitions and (i) and (iv) represent allowed transitions. 
In order to calculate decay rates of hindered transitions we need to include relativistic corrections.There are three main types of corrections: relativistic modification of the non relativistic wave functions, relativistic modification of the electromagnetic transition operator, and finite-size corrections. In addition to these there are additional corrections arising from the quark
anomalous magnetic moment. Corrections to the wave function that give contributions to the transition amplitude are of two categories: \\
1) higher order potential corrections, which are distinguished as a) Zero recoil effect and b) recoil effects of the final state meson and 2)Colour octet effects. The colour octet effects are not included in potential model formulation and are not considered so far in radiative transitions.
\begin{table*}
\centering
\caption{\label{E1}\bf E1 transition rates of $B_c$ meson.}
\begin{tabular*}{\textwidth}{@{\extracolsep{\fill}}lrrrrrrrrl@{}}
\hline
Transition&\multicolumn{1}{c}{k$_0$}&\multicolumn{1}{c}{This work}&\multicolumn{1}{c}{Ref. \cite{DR03}}&\multicolumn{1}{c}{Ref. \cite{EC94}}&\multicolumn{1}{c}{Ref. \cite{VA95}}&\multicolumn{1}{c}{Ref.\cite{FL}}\\
&MeV&keV&keV&keV&keV&keV\\
\hline
$1^3P_0\rightarrow 1^3S_1\gamma$&348.527&42.384&75.5&79.2&65.3&74.2\\
$1P1\rightarrow 1^3S_1\gamma$&437.575&83.879&87.1&99.5&77.8&75.8\\
$1P1'\rightarrow 1^3S_1\gamma$&494.615&121.143&13.7&0.1&8.1&26.2\\
$1^3P_2\rightarrow 1^3S_1\gamma$&446.371&89.04&122&112.6&102.9&126\\
$1P1\rightarrow 1^1S_0\gamma$&473.213&106.088&18.4&0&11.6&32.5\\
$1P1'\rightarrow 1^1S_0\gamma$&529.934&148.992&147&56.4&131.1&128\\
$2^3S_1\rightarrow 1^3P_0\gamma$&175.953&3.635&5.53&7.8&7.7&9.6\\
$2^3S_1\rightarrow 1P1\gamma$&83.181&0.384&7.65&14.5&12.8&13.3\\
$2^3S_1\rightarrow 1P1'\gamma$&22.416&0.00751&0.74&0&1.0&2.5\\
$2^3S_1\rightarrow 1^3P_2\gamma$&73.879&0.269&7.59&17.7&14.8&14.5\\
$2^1S_0\rightarrow 1P1\gamma$&70.979&0.238&1.05&0&1.9&6.4\\
$2^1S_0\rightarrow 1P1'\gamma$&10.104&0.00068&4.40&5.2&15.9&13.1\\
$2^3P_0\rightarrow 1^3S_1\gamma$&573.927&0&&21.9&16.1\\
$2P1\rightarrow 1^3S_1\gamma$&88.815&0&&22.1&15.3\\
$2P1'\rightarrow 1^3S_1\gamma$&938.854&0&&2.1&2.5\\
$2^3P_2\rightarrow 1^3S_1\gamma$&859.837&0&&25.8&19.2\\
$2P1\rightarrow 1^1S_0\gamma$&917.035&0&& &3.1\\
$2P1'\rightarrow 1^1S_0\gamma$&971.788&0&& &20.1\\
$2^3P_0\rightarrow 2^3S_1\gamma$&63.253&0.422&34.0&41.2&25.5\\

$2P1\rightarrow 2^3S_1\gamma$&397.439&104.751&45.3&54.3&32.1\\
$2P1'\rightarrow 2^3S_1\gamma$&456.65&158.896&10.4&5.4&5.9\\
$2^3P_2\rightarrow 2^3S_1\gamma$&371.628&85.639&75.3&73.8&49.4\\
$2P1\rightarrow 2^1S_0\gamma$&409.075&114.223&13.8& &8.1\\
$2P1'\rightarrow 2^1S_0\gamma$&468.191&171.244&90.5&&58.0\\
\hline
\end{tabular*}
\end{table*}
The spherical Bessel function $j_0(kr/2)$ introduced in equation (\ref{decay}) takes into account the so called finite-size effect (equivalently, re summing multipole-expanded magnetic amplitude to all orders). For small $k$, $j_0(kr/2)\rightarrow 1$, so that transitions with $n'=n$ have favoured matrix elements, though the corresponding partial decay widths are suppressed by smaller
$k^3$ factors. For large value of photon energy ($k$) transitions with $n\neq n'$ have favoured the matrix element, since $j_0(kr/2)$ becomes very small. $M1$ transition rates are very sensitive to hyperfine splitting of the levels due to the $k^3$ factor in equation (\ref{decay}).
     
There have been many models which study the radiative decays of $B_c$ meson using non relativistic and relativistic quark models. Eichten and Quigg \cite{EC94} calculated the radiative M1 transition rates for the allowed and hindered transitions. They used the equation (\ref{decay}) in their potential model approach to determine the M1 transition rates of $B_c$ meson. Allowed transition rates for processes (i) and (iv) were found to be 0.0040 keV and 0.130 keV respectively. Hindered transition rates for the processes (ii) and (iii) were 0.253 keV and 0.223 keV respectively. Abd El-Hady \textit{et al} \cite{AA05} have investigated the radiative decay properties of $B_c$ meson in a Bethe-Salpeter model. The allowed transition rates for processes (i) and (iii) were found to be 0.0037 keV and 0.0189 keV respectively. The hindered transition rates for the processes (ii) and (iv) were found to be 0.135 keV and 0.1638 respectively. Ebert \textit{et al} \cite{DR03} have studied these M1 transitions including full relativistic corrections in their relativistic model. They depend explicitly on the Lorentz structure of the non relativistic potential. Several sources of uncertainty make M1 transitions particularly difficult to calculate. The leading-order results depend explicitly on the constituent quark masses, and corrections depend on the Lorentz structure of the potential. They estimated the allowed transition rates to be 0.033 keV and 0.017 keV respectively. For the hindered transition, decay rates were found to be 0.428 keV and 0.488 keV.  Also it is clear from their calculations that the predicted decay rates for hindered transitions which are increased by relativistic effects almost by a factor of 3 and they are larger than the rates of allowed M1 transitions by an order of magnitude.

 We have calculated the M1 transition rates for $c\bar{b}$ meson states using equation (\ref{decay}).
The resulting M1 radiative transition rates of these states are presented in table \ref{M1}. In this table we give calculated values for decay rates of M1 radiative transition in comparison with the other relativistic and non relativistic quark models. We see from these results that the relativistic effects play a very important role in determining the $B_c$ meson M1 transition rates. The relativistic effects reduce the decay rates of allowed transitions and increase the rates of hindered transitions. The M1 transition rates calculated in our model agree well with the values predicted by other theoretical models.  
 \begin{table*}[h]
\centering
\caption{\label{M1}\bf M1 transition rates for the $B_c$ meson.}
\setlength{\tabcolsep}{2pt}
\begin{tabular}{cccccccccc}
\hline
Transition&$k_{0}(MeV)$&$\Gamma(keV)$& $\Gamma(keV)$  &$\Gamma(keV)$&$\Gamma(keV)$&$\Gamma(keV)$&$\Gamma(keV)$ &$\Gamma(keV)$\\
&&This work&Ref.\cite{AA05}&Ref.\cite{DR03}&Ref.\cite{FL}&Ref.\cite{SA95}&Ref.\cite{DR03}&Ref.\cite{EC94}&\\
\hline
$1~ ^3S_1\rightarrow 1 ^1S_0\gamma$&38.193&0.0185&0.0189&0.033&0.059&0.060&0.073&0.135\\
$2~ ^3S_1\rightarrow 2 ^1S_0\gamma$&12.329&0.0018&0.0037&0.017&0.012&0.010&0.030&0.029\\
$2~ ^3S_1\rightarrow 1 ^1S_0\gamma$&550.614&0.193&0.1357&0.428&0.122&0.098&0.141&0.123\\
$2~ ^3S_1\rightarrow 1 ^3S_1\gamma$&515.410&0.123&0.1638&0.488&0.139&0.096&0.160&0.093\\
\hline
\end{tabular}
\end{table*}

\subsection{Weak Decays and Life Time of $B_c$ meson} 
 
In accordance with the classification given in section \ref{weak}, the total decay width can be written as the sum over partial widths
 \begin{equation}
 \Gamma(B_c\rightarrow X)=\Gamma(b\rightarrow X)+\Gamma(c\rightarrow X)+\Gamma(ann)
 \end{equation}
 In the spectator approximation:
 \begin{equation}
 \Gamma_1(\bar{b}\rightarrow X)=\frac{9G^2_F|V_{cb}|^2m^5_b}{192\pi^3}\label{eq1}
 \end{equation}
 Calculated value of $\Gamma_1(\bar{b}\rightarrow X)$ is $1.041\times 10^{-3}~\rm{eV}$ 
 and
\begin{equation}
 \Gamma_2(c\rightarrow X)=\frac{5G^2_F|V_{cs}|^2m^5_c}{192\pi^3} \label{eq2}
 \end{equation}
 Calculated value of $\Gamma_2(c\rightarrow X)$ is $8.958\times 10^{-4}~\rm{eV}$.\\
 In the above expressions $V_{cb}$ and $V_{cs}$ are the elements of the CKM matrix, $G_F=1.16637\times 10^{-5}$ is the Fermi coupling constant, $m_c$ and $m_b$ are the masses of c and b quarks respectively. The decay widths are calculated using $|V_{bc}|=0.044$ \cite{PDG} and $|V_{cs}|=0.975$ \cite{PDG}.\\

The leptonic partial widths are probe of the compactness of quarkonium system and provide important information complementary to level spacings. The quark-antiquark assignments for the vector mesons, as well as the fractional values for the quark charges, are tested from the values of their leptonic decay widths. The decay of vector meson into charged leptons proceeds through the virtual photon $(q\bar{q}\rightarrow l^+l^-)$. The $^3S_1$ and $^3D_1$ states have quantum numbers of a virtual photon, $J^{PC}=1^{--}$ and can annihilate into lepton pairs through one photon.
Annihilation widths such as $c\bar{b}\rightarrow l\nu_l$ are given by the expression
\begin{equation}
\Gamma_3=\frac{G^2_F}{8\pi}|V_{bc}|^2f^2_{B_c}M_{B_c}\sum_i m^2_i\left(1-\frac{m^2_i}{M^2_{B_c}}\right)C_i\label{eq3}
\end{equation}
Calculated value of $\Gamma_3$ is $5.663\times 10^{-6}~\rm{eV}$.\\
Here $m_i$ is the mass of the heavier fermion in the given decay channel. For lepton channels $C_i=1$ while for quark channels $C_i=3|V_{q\bar{q}}|^2$ and $f_{B_c}$ is the pseudo scalar decay constant for $B_c$ meson. \\

 Adding these results we get the total decay width $\Gamma(\rm{total})=\Gamma_1+\Gamma_2+\Gamma_3=19.428\times 10^{-4}~\rm{eV}$ corresponding to a life time of $\tau=0.339~\rm{ps}$.\\

The pseudo scalar decay constant $f_{B_c}$ is defined by:
\begin{equation}
\bra{0}\bar{b}(x)\gamma^\mu c(x)\ket{B_c(k)}=if_{B_c}V_{cb}k^\mu 
\end{equation}  
 where $k^\mu$ is the four-momentum of the $B_c$ meson. In the non relativistic limit the pseudo scalar decay constant is proportional to the wave function at the origin and is given by van Royen-Weisskopf formula \cite{RV67}
 \begin{equation}
 f_{B_c}=\sqrt{\frac{12}{M_{B_c}}}\psi(0)
 \end{equation}
 Here $\psi(0)$ is wavefunction at the origin.
The values of decay constant in various theoretical models are listed in table \ref{fbc} and in table \ref{T1} we compare the life time of $B_c$ meson calculated in our model with other models.
\begin{table}[h]
\centering
\caption{\label{fbc}\bf Comparison of predictions for the pseudo scalar decay constant of the $B_c$ meson ($f_{B_c}$).}
\setlength{\tabcolsep}{2pt}
\begin{tabular}{ccccccc}
\hline
This work&Ref.\cite{WS81}&Ref. \cite{AM80}&Ref.\cite{CJ77}&Ref.\cite{CT96}\\
\hline
554.125&500&512&479&440$\pm$20\\
\hline
\end{tabular}
\end{table}
\begin{table*}[h]
\centering
\caption{\label{T1}\bf Comparison of life time of $B_c$ meson (in ps).}
\setlength{\tabcolsep}{2pt}
\begin{tabular}{ccccccc}
\hline
This work&Experiment\cite{PDG}&Ref.\cite{AA99}&Ref.\cite{VA95} &Ref.\cite{KVV}&Ref. \cite{GS85}\\
\hline
0.339&0.452$\pm 0.033$&0.47&0.55$\pm 0.15$&0.50&0.75\\
\hline
\end{tabular}
\end{table*} 
\subsection{Strong Decays}
The $c\bar{b}$ states which lie below BD threshold are stable against strong decays. However, the states which are above the BD threshold undergo two body strong decays. We have calculated strong decay widths of $c\bar{b}$ states which lie above the BD threshold using the equation (\ref{strong}). The decay widths are calculated within the $^3P_0$ pair creation model. The results are presented in table \ref{strong1}.
\begin{table*}[h]
\centering
\caption{\label{strong1}\bf Strong decay widths of the $B_c$ meson.}
\setlength{\tabcolsep}{2pt}
\begin{tabular}{ccccccccc}
\hline
Transition&$\Gamma(MeV)$\\
\hline
$2~ ^1P_1\rightarrow B^*+D$&54.599\\
$2~ ^3P_1\rightarrow B^{*}+D$&2.145\\
$2~ ^3P_2\rightarrow B+D$&99.386\\
$2~ ^3P_2\rightarrow B^0+D^0$&108.185\\
$2~ ^3P_2\rightarrow B^*+D$&31.247\\
$1~^3D_2\to B^*+D$&0.198\\
$1~^3D_2\to B^*_s+D_s$&5.837\\
$1~^3D_2\to B^*+D^*$&2.123\\
$1~^3D_2\to B^*_s+D^*_s$&20.885\\
\hline
\end{tabular}
\end{table*}
\section{Conclusions}
\label{sec:C}
The complete spectrum of $c\bar{b}$ states has been calculated in a relativistic quark model with coupled channel effects. We have calculated the meson loop effects on the masses of 1S, 2S, 1P, 2P and 1D $c\bar{b}$ states. The mass shifts calculated due to these loop effects are large.  The ground state mass of $c\bar{b}$ state  calculated in our model matches the experimental data. When the results for $c\bar{b}$ state mass spectrum are compared with the previous calculations, it is found that the predictions for the mass spectrum agree within a few MeV. The differences between the predictions in most cases do not exceed 30 MeV and the higher orbitally excited states are 50-100 MeV heavier in our model. The hyperfine splitting of the ground state  vector and pseudo scalar $c\bar{b}$ states in our model is in good agreement with the prediction made by other theoretical models. The ground state pseudo scalar $B_c$ and vector $B^*_c$ meson masses lie within the ranges quoted by Kwong and Rosner in their survey of techniques for estimating the masses of the $c\bar{b}$ ground state: i.e., $6194~ \rm{MeV}<M_{B_c}<6292~\rm{MeV}$ and $6284~\rm{MeV}<M_{B^*_c}<6357~\rm{MeV}$. \\

 The difference of quark flavours forbids $B_c$ meson from annihilation into gluons. As a result the excited $B_c$ meson states lying below the BD production threshold (i.e. with $M<M_D+M_B=7143.1\pm 2.1$ GeV) undergo radiative transition to ground state which then decays weakly. Radiative decays are the dominant decay modes of the $B_c$ excited states having widths of about a fraction of MeV. Therefore, it is very important to determine the masses and the radiative decay widths of $B_c$ meson accurately in order to understand the $B_c$ spectrum and distinguishing exotic states. The radiative E1 and M1 decay rates of $c\bar{b}$ states have been calculated using spectroscopic parameters obtained from RQM. Most of our predictions for the E1 decay rates are in good agreement with the other theoretical calculations. The differences in the prediction for the decay rates in various theoretical models can be attributed to the  differences in mass predictions, wave function effects and singlet - triplet mixing angels. The calculated  M1 transition rates reasonably agree with the other theoretical model predictions as listed in table \ref{M1}. It is clearly seen in this calculation that the relativistic effects play an important role in determining the radiative transition rates, since the hindered transition rates are suppressed due to the wave function orthogonality in the NRQM formalism. The inclusion of these relativistic effects enhances the hindered transition rates and reduces the allowed transition rates. It is evident from the table that the hindered transition rates are larger than the allowed transition rates by an order of magnitude. Experimental results for the masses of excited states and radiative decays of $B_c$ meson are needed to clarify these predictions. Experimental results will give us more insight into $B_c$ spectrum and will help us to clarify the hyperfine splitting calculated in different models.\\
 
We have done an estimation of weak decay widths in the spectator quark approximation and calculated the life time of $c\bar{b}$ state. We get about 53\% branching ratio for $b$-quark decays, about 46\% for $c$-quark decays and about 1\% branching ratio in annihilation channel. The life time of $c\bar{b}$ state predicted in this calculation is listed in table \ref{T1} and is found to be in good agreement with experimental value as well as with other theoretical predictions. The decay constant of $c\bar{b}$ state ($f_{B_c}$) has been calculated and compared with other model predictions and it is found that the decay constant is consistent with these predictions.  We have calculated two body strong decay widths of $c\bar{b}$ states in the framework of $^3P_0$ pair creation model.\\

A simple relativistic model employing COGEP and
harmonic-oscillator confinement potential along with coupled channel effects used in this study is successful to predict the various properties of $c\bar{b}$ states and this can shed
further light on their non leptonic transitions.

\begin{center}
\textbf{Acknowledgements}
\end{center}
One of the authors (APM) is grateful to BRNS,                                                             DAE, India for granting the project and JRF (37(3)/14/21/2014BRNS). 

%\section*{References}
\bibliography{mybib}

\end{document}